\documentclass[epj]{svjour}
\usepackage{graphics}
\usepackage{graphicx}
\usepackage{psfrag}
\begin{document}
\title{Infrared Gluon and Ghost Propagators from Lattice QCD}
\subtitle{Results from large asymmetric lattices}
\author{O. Oliveira \and P. J. Silva 
}                     
%
%
\institute{Centro de F\'{\i}sica Computacional, Universidade de Coimbra, 
           3004 516 Coimbra, Portugal}
\date{Received: date / Revised version: date}
%
\abstract{We report on the infrared limit of the quenched lattice Landau gauge
gluon and ghost propagators as well as the strong coupling constant computed
from large asymmetric lattices. The infrared lattice propagators
are compared with the pure power law solutions from Dyson-Schwinger equations
(DSE). For the gluon propagator, the lattice data is compatible with the DSE 
solution. The preferred measured gluon exponent being $\sim 0.52$, favouring
a vanishing propagator at zero momentum.
The lattice ghost propagator shows finite volume effects and, for the
volumes considered, the propagator does not follow a pure power law.
Furthermore, the strong coupling constant is computed and its infrared 
behaviour investigated.
\PACS{{12.38.-t}{Quantum chromodynamics} \and
      {11.15.Ha}{Lattice gauge theory} \and
      {12.38.Gc}{Lattice QCD calculations} \and
      {12.38.Aw}{General properties of QCD} \and
      {14.70.Dj}{Gluons} \and
      {14.80.-j}{Other particles}
     } 
} 
\maketitle

\section{Introduction}
\label{intro}
\begin{figure}
\psfrag{EIXOX}{\Huge k (GeV)}
\psfrag{EIXOY}{\Huge $Z_{gluon} ( k^2 )$}
\psfrag{GEIXY}{\Huge $Z_{ghost} ( k^2 )$}
\resizebox{0.75\textwidth}{!}{\includegraphics[angle=-90,scale=0.006]{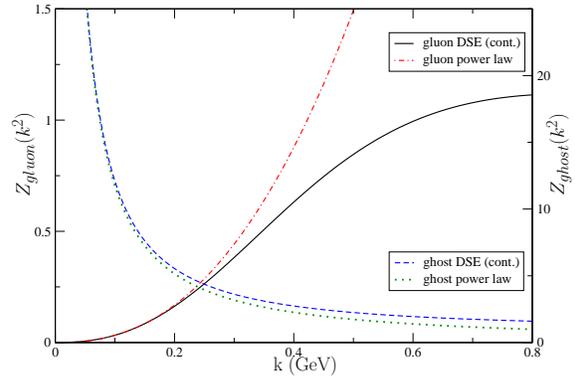}
}
\caption{DSE gluon and ghost dressing functions versus the pure power law 
solutions. The data is taken from \cite{FisPen06}.}
\label{DSEfig}
\end{figure}

In the pure gauge SU(3) Yang-Mills theory, a number of authors has been using 
first principles approa\-ches, i.e. Dyson-Schwinger equations (DSE) and 
lattice QCD methods, to investigate the infrared gluon and ghost propagators 
in Landau gauge, respectively,
\begin{eqnarray}
  D^{ab}_{\mu\nu} (k) & = & \delta^{ab}  \,
      \left( \delta_{\mu\nu} - \frac{k_\mu k_\nu}{k^2} \right) \, 
  D (k^2) \, , \\
  G^{ab}_{\mu\nu} (k) & = & - \, \delta^{ab}  \,
                                  G (k^2) \, ,
\end{eqnarray}
and the strong coupling constant \cite{Prosperi06} defined as
\begin{equation}
  \alpha_S ( k^2 ) ~ = ~ \alpha_S ( \mu^2 ) \, Z^2_{ghost} ( k^2 ) \,
                                               Z_{gluon} ( k^2 ) \, ;
\label{alfaS}
\end{equation}
$Z_{ghost} ( k^2 ) = k^2 G( k^2 )$ and $Z_{gluon} ( k^2 ) = k^2 D( k^2 )$ are
the ghost and gluon dressing functions. In part, these studies 
have been trigerred by the solution of the DSE \cite{smekal97} which assuming 
infrared ghost dominance and infrared finiteness of the loop-integrals predicts
pure power laws for the dressing functions
\begin{equation}
 Z_{gluon} ( k^2 ) ~ = ~ A \left( k^2 \right)^{\kappa'} \, , 
 Z_{ghost} ( k^2 ) ~ = ~ B \left( k^2 \right)^{- \kappa} \, .
\label{DSEsol}
\end{equation}
Moreover, the DSE solution relates the exponents of the two propagators,
$\kappa' = 2 \kappa$, and 
predicts a finite strong coupling constant at zero momentum. The infrared
solution predicts a vanishing gluon propagator and an infinite ghost propagator
at zero momentum ($\kappa > 0.5$). The infrared behaviour of the propagators 
can be related to gluon confinement criteria \cite{Fischer06}. 
Looking at this particular DSE solution for the gluon propagator, the 
comparison with the pure power law, see figure \ref{DSEfig}, shows that 
the pure power law is valid only for momenta $k < 200$ MeV. Notice that for
the ghost, the deviations from a pure power law start earlier.

One should have in mind that the solution discussed above is a particular
solution of the DSE. Indeed, there are, in the literature, different types of 
solutions for the DSE \cite{Aguillar}. Moreover, the authors of
\cite{franceses} have investigated a generalization of the conditions assumed
in \cite{smekal97} and found alternative behaviours for the infrared 
gluon and ghost propagators. Given the different scenarios,
it would be important if lattice QCD can provide additional input, helping
to check if any of the proposed solutions reproduces the lattice data. 
For a recent discussion on the infrared behaviour of the gluon and ghost
propagators see \cite{FischerPaw06}.

For the lattice it is a challenge to investigate such low momenta. A possible 
way out is to consider large asymmetric lattices, which allows to access the
momenta required in such an investigation 
\cite{oliveira05,silva05,oliveira06,silva06prd}. Of course, there are lattice
effects that have to be carefully estimated.
Here we report on the gluon, ghost and strong coupling constant computed from
large asymmetric lattices.

\section{Gluon Propagator}
\label{sec:1}

In \cite{silva06prd} we have investigated the infrared lattice Landau gauge 
gluon propagator. For notation and definitions see the above cited article.
The simulation uses the Wilson action with $\beta = 6.0$ and
combines a number of asymmetric lattices $L^3 \times 256$, 
$L = 8, 10, 12, 14, 16, 18$ to allow for $L \rightarrow + \infty$ 
extrapolation. In what concerns the time direction, the $16^3 \times 256$ and 
$16^3 \times 128$ gluon propagators were compared and no deviations were 
observed. This suggests that we have a sufficient number of points in the
time direction.

For the lattices with the largest time direction, the minimum momentum being 
accessed is 47 MeV, while for the $16^3 \times 128$ the minimum momentum is 
98 MeV. For the largest lattices with $T = 256$ a pure power law in the 
infrared is observed. In contrast, for the lattice of $16^3 \times 128$,
power law fits are very poor with $\chi^2/d.o.f > 10$. This result 
is not surprising, given the smallest range of momenta available for this 
lattice (97 - 294 MeV) and the validity of the pure power law. Nevertheless, 
the $16^3 \times 128$ will allow us to estimate Gribov copies effects, with
reasonable computational effort.

Let us summarize the results reported in \cite{silva06prd} using the 
simulations with the largest time extension. Asymmetric lattices display
finite size effects (see figures 4 and 5 of \cite{silva06prd}). However, the
extrapolation towards $L \rightarrow + \infty$ is smooth. The extrapolated 
gluon propagator was well reproduced by a pure power law, with measured
$\kappa = 0.49 - 0.52$. Clearly, the lattice data favors $\kappa \sim 0.52$,
in agreement with other theoretical estimates. Unfortunately, one cannot yet
provide a definitive answer concerning the value of the gluon 
propagator at zero momentum. We are currently engaged in trying to give such
an answer.

\begin{figure}
\resizebox{0.75\textwidth}{!}{\includegraphics[scale=0.006]{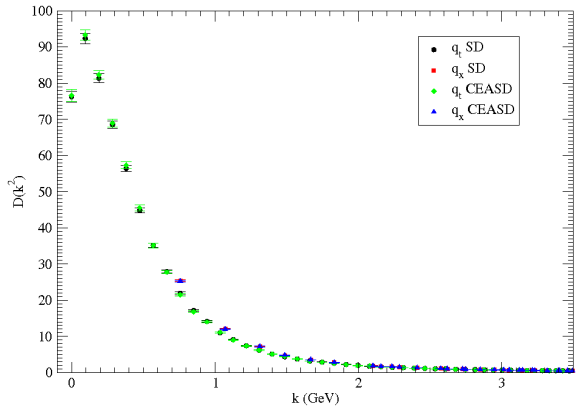}
}
\caption{Bare gluon propagator for $16^3 \times 128$.}
\label{gluongribov}
\end{figure}

For the $16^3 \times 128$ lattice, Gribov copies effects were estimated by
comparing the gluon propagator computed from 164 configurations gauge fixed as
in \cite{silva06prd} (SD in the figures) and the same gauge configurations 
gauge fixed using the method described in \cite{oliveira04} (CEASD in the 
figures), which aims to estimate the absolute maximum of the gauge fixing 
function. The gluon propagator given by the two methods is, within the 
statistical precision of the simulation, the same (see figure 
\ref{gluongribov}), with the various fits to the lattice data
(infrared, ultraviolet, all momenta) reproducing similar results. Therefore, 
no visible effects of Gribov copies are observed on the gluon propagator.

\section{Ghost Propagator}

The ghost propagator \cite{ghost} was 
computed using two different methods \cite{ss,cuco}. The first allows, by 
solving a linear 
system, to access all momenta. Its drawback
being that, by computing $G(x,0)$, the statistical errors in the propagator 
are much larger. The second method requires an inversion of the same linear
system per momentum, which is more computationally demanding. However, 
for the momenta considered, it determines the propagator with better 
statistical accuracy. 
Computing the propagator using two methods provides a valuable
cross-check.

In the following, for the ghost propagator the colour average was always
performed. For the propagator computed with the \cite{ss} method, in order
to reduce the statistical error, seven different sources for the linear system
were considered and their result averaged.

\begin{figure}
\resizebox{0.75\textwidth}{!}{\includegraphics[scale=0.006]
{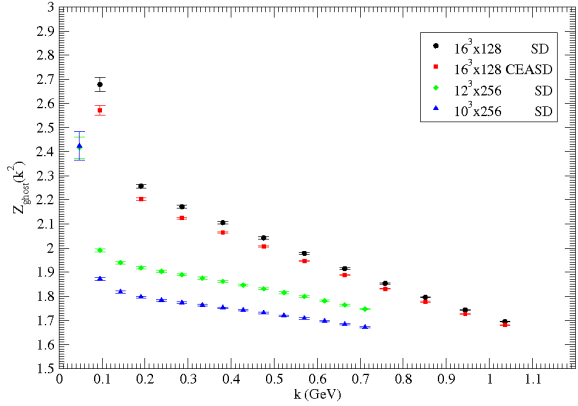}
}
\caption{Bare ghost dressing functions for a $16^3 \times 128$ and 
$L^3 \times 256$, $L = 10$, $12$ lattices.}
\label{ghostgribov}
\end{figure}

For the ghost propagator, one observes sizeable finite size effects (see 
figure \ref{ghostgribov}), in agreement with the SU(2) study \cite{cuco06}, 
and a very sign of clear Gribov copies effects for a large range of momenta. 

In what concerns finite volume effects, figure \ref{ghostgribov} shows that
the ghost dressing function is enhanced when the lattice volume is increased. 
Moreover, the larger volumes have a larger dressing function and,
for the smallest momenta, the derivatives of the dressing function seem to
become smaller when the volume increases.
The effect of Gribov copies, seen in figure \ref{ghostgribov}, cannot be 
eliminated by the renormalization of the propagator.

To study the compatibility of the ghost propagator with the DSE solutions,
the lattice data was fitted to pure power laws and na\"{\i}ve corrections to 
the
pure power law. It turns out that the data are not described by any of the
functions considered. Given the results of a similar study for the
$16^3 \times 128$ gluon propagator, this result is not a complete surprise.

\section{Strong Coupling Constant}

\begin{figure}
\psfrag{EIXOY}{\Huge $\alpha ( k^2 ) / \alpha ( \mu^2 )$}
\psfrag{EIXOX}{\Huge k (GeV)}
\resizebox{0.75\textwidth}{!}{\includegraphics[angle=-90,scale=0.006]
{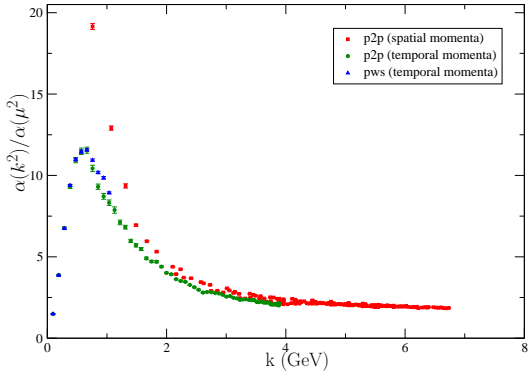}
}
\caption{Strong coupling constant for $16^3 \times 128$
configurations for the CEASD gauge fixing method; ``p2p'' (``pws'') stands
for the ghost computed using \cite{ss} (\cite{cuco}) method.}
\label{FalfaS1}
\end{figure}

\begin{figure}
\psfrag{EIXOY}{\Huge $\alpha ( k^2 ) / \alpha ( \mu^2 )$}
\psfrag{EIXOX}{\Huge k (GeV)}
\resizebox{0.75\textwidth}{!}{\includegraphics[angle=-90,scale=0.006]
{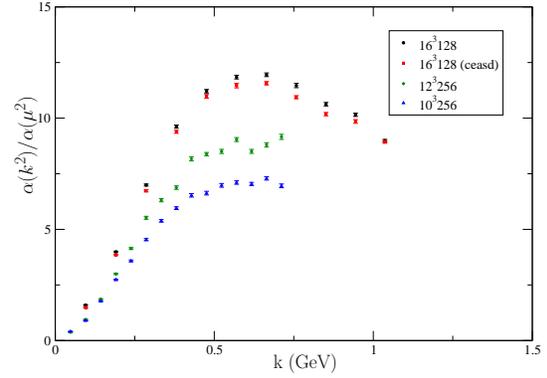}
}
\caption{Small momenta strong coupling for all the simulations.}
\label{FalfaS2}
\end{figure}

The strong coupling constant as defined by equation (\ref{alfaS}) is displayed
in figure \ref{FalfaS1} for the $16^3 \times 128$ lattice and CEASD gauge 
fixing method. For the other simulations and the SD method, the measured
coupling constants are, qualitatively, similar. Figure \ref{FalfaS2} show
the strong coupling constant for small momenta, for all the simulations 
considered previously.

The simulations show a decreasing lattice strong coupling constant for small
momenta, in agreement with previous lattice calculations 
\cite{Furui,Sternbeck}. From our simulation, it is not clear if 
$\alpha_S (0)$ is vanishing. We have tried a number of fits to the 
strong coupling constant as a function of the momentum, some requiring 
$\alpha_S (0) = 0$ and some leaving $\alpha_S (0)$ as a free parameter, and the
data can be 
equally described by the two situations. For small momenta, only for 
the $16^3 \times 128$ data and the CEASD gauge fixing algorithm, is $\alpha_S$
described by a pure power law $\alpha_S ( q^2 ) = ( q^2 )^{\kappa_\alpha}$, 
with $\kappa_\alpha = 0.69$ suggesting a vanishing strong coupling constant
at zero momentum. 
However, notice that the data show that $\alpha_S$ increases with 
lattice volume as seen in figure \ref{FalfaS2} and is affected by Gribov
copies. Further studies will hopefully lead to definitive conclusions.

%


\begin{thebibliography}{}
\bibitem{ghost}
The definitions concerning the ghost propagator can be found in
the articles where the lattice ghost propagator has been computed so far.
\end{thebibliography}
\end{document}